\documentclass[doublecol]{epl2} 
\usepackage{graphicx}
\usepackage{amsmath,amssymb,revsymb}
\usepackage{color}

\title{Magnetic field reversals in an experimental turbulent dynamo}
\author{M. Berhanu\inst{1}, R. Monchaux\inst{2}, S. Fauve\inst{1}, N. Mordant\inst{1}, F. P\'etr\'elis\inst{1}, A. Chiffaudel\inst{2}, F. Daviaud\inst{2} B. Dubrulle\inst{2}, L. Mari\'e\inst{2,+}, F. Ravelet\inst{2,*}, M. Bourgoin\inst{3,\#}, 
Ph. Odier\inst{3}, J.-F. Pinton\inst{3}, R. Volk\inst{3} }
\shortauthor{Berhanu \etal}
\institute{
\inst{1} Laboratoire de Physique Statistique, \'Ecole Normale Sup\'erieure CNRS UMR8550, 24 rue Lhomond, F-75005 Paris (France)\\
\inst{2} Service de Physique de l'\'Etat Condens\'e, Direction des Sciences de la Mati\`ere\\ CNRS URA 2464, CEA-Saclay, F-91191 Gif-sur-Yvette (France)\\
\inst{3} Laboratoire de Physique, \'Ecole Normale Sup\'erieure de Lyon,  CNRS UMR5672, 46 all\'ee dÕItalie, F-69364 Lyon (France)
}


\abstract{We report the first experimental observation of reversals of a dynamo field generated in a laboratory experiment based on a turbulent flow of liquid sodium. The magnetic field randomly switches between two symmetric solutions ${\mathbf B}$ and $-{\mathbf B}$. We observe a hierarchy of time scales similar to the Earth's magnetic field: the duration of the steady phases is widely distributed, but is always much longer than the time needed to switch polarity. In addition to reversals we report excursions. Both coincide with minima of the mechanical power driving the flow. Small changes in the flow driving parameters also reveal a large variety of dynamo regimes.}

\begin{document}

\maketitle

Dynamo action is the instability mechanism by which mechanical energy is partially converted into magnetic energy by the motion of an electrically conducting fluid~\cite{moffatt}. It is believed to be at the origin of the magnetic fields of planets and most astrophysical objects.  One of the most striking features of the Earth's dynamo,  revealed by paleomagnetic studies~\cite{valet}, is the observation of irregular reversals of the polarity of its dipole field. This behaviour is allowed from the constitutive equations of magnetohydrodynamics~\cite{moffatt} and has been observed in numerical models~\cite{numterre}. On the other hand, industrial dynamos routinely generate currents and magnetic fields from mechanical motions. In these devices, pioneered by Siemens~\cite{siemens}, the path of the electrical currents and the geometry of the (solid) rotors are completely prescribed. As it cannot be the case for planets and stars, experiments aimed at studying dynamos in the laboratory have evolved towards relaxing these constraints. Solid rotor experiments~\cite{lowes} showed that a dynamo state could be reached with prescribed motions but currents free to self-organize. A landmark was reached in 2000, when the experiments in Riga~\cite{gailitis} and Karlsruhe~\cite{stieglitz} showed that fluid dynamos could be generated by organizing favourable sodium flows, the electrical currents being again free to self-organize. For these experiments, the self-sustained dynamo fields had simple time dynamics (a steady field in Karlsruhe and an oscillatory field in Riga). No further dynamical evolution was observed. The search for more complex dynamics, such as exhibited by natural objects, has motivated most teams working on the dynamo problem to design experiments with less constrained flows and a higher level of turbulence~\cite{nondyn}. The von K\'arm\'an sodium experiment (VKS) is one of them. It has recently shown regimes where a statistically stationary dynamo self-generates~\cite{monchaux}. We report here the existence of other dynamical regimes and describe below the occurence of irregular reversals and excursions.

\begin{figure}[h!]
\centerline{\includegraphics[width=8cm]{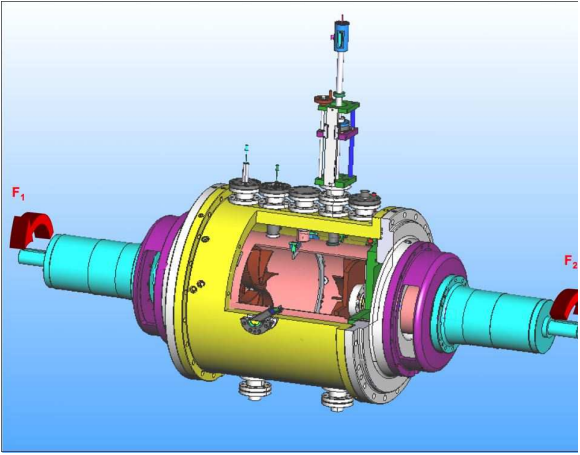}}
\caption{The VKS2 set-up is designed to generate a dynamo flow in an electrically conducting fluid. The overall vessel is a copper cylinder of radius 289 mm and length 604 mm. The flow itself is confined within an inner copper cylinder (radius $ R=206$~mm, length 524 mm, thickness 5 mm), with sodium at rest between the inner and outer cylinders. An annulus of inner radius 175 mm (thickness 5 mm) is fixed along the inner cylinder in the mid-plane between the disks. The counter-rotating iron impellers have radius 154.5 mm and are set 371 mm apart in the inner vessel; they are fitted with 8 curved blades of height $h=41.2$~mm. Their rotation frequencies are independently adjustable, up to 26 Hz.  Magnetic measurements are made using a temperature controlled, 3D Hall probe mounted flush on the flow boundary, at the inner cylinder.} 
\label{setup}
\end{figure}

Magnetic induction properties of turbulent swirling flows (VKS and others) have been widely studied experimentally~\cite{nondyn,vks}. Dynamo action by these flows has also received strong support from numerical simulations~\cite{numer,ravelet}. The VKS experimental set-up, shown in Figure~1, has been described in~\cite{monchaux}. The fluid is liquid sodium, chosen for its high electrical conductivity ($\sigma = 10^7 \; {\rm ohm}^{-1}{\rm m}^{-1}$), low density ($\rho = 930 \; {\rm kg} \, {\rm m}^{-3}$) and low melting point ($98^{\circ }$C). The net volume of sodium is roughly 160 L.  A turbulent von K\'arm\'an flow is generated by two counter-rotating iron impellers (rotation frequencies $F_1$ and $F_2$). Its mean structure has the following characteristics: the fluid is ejected radially from the disks by centrifugal force and loops back towards the axis in the mid-plane between the impellers. A strong differential rotation is superimposed on this poloidal flow, which generates a high shear in the mid-plane.  The flow maximum driving power is 300 kW, and cooling is performed using an oil flow inside the copper walls of the vessel. It allows experimental runs at constant temperatures between $110^{\circ}$C and $160^{\circ}$C. The integral Reynolds numbers are defined as ${\rm Re_{\it i}} = 2\pi K R^2 F_i/\nu$ and take values up to $5\, 10^6$ where  $\nu$ is the fluid viscosity and $K=0.6$ is a coefficient that measures the efficiency of the driving impellers~\cite{ravelet}. Corresponding magnetic Reynolds numbers, ${\rm R_{m{\it i}}} = 2 \pi K \mu_0 \sigma R^2 F_i$, up to 49 at $120^{\circ}$C are reached -- $\mu_0$ is the magnetic permeability of vacuum. The magnetic field is measured with local Hall probes inserted inside the fluid.\\

When the impellers are operated at equal and opposite rotation rates $F$, a fully turbulent dynamo is observed when $F$ is larger than about 17 Hz (${\rm R_m}=31$)~\cite{monchaux}. The self-sustained magnetic field is statistically stationary with either polarity in this case. In the experiment, the rotation rates $(F_1, F_2)$ of the driving impellers can be independently adjusted and this gives an additional degree of freedom. Starting from a symmetric flow forcing, $F_1=F_2$, one can progressively change the rotation frequency of one disk and explore regimes in which the faster disk imposes some kind of global rotation to the flow, a feature common to most natural dynamos. 

\begin{figure}[h!]
\centerline{\includegraphics[width=11cm]{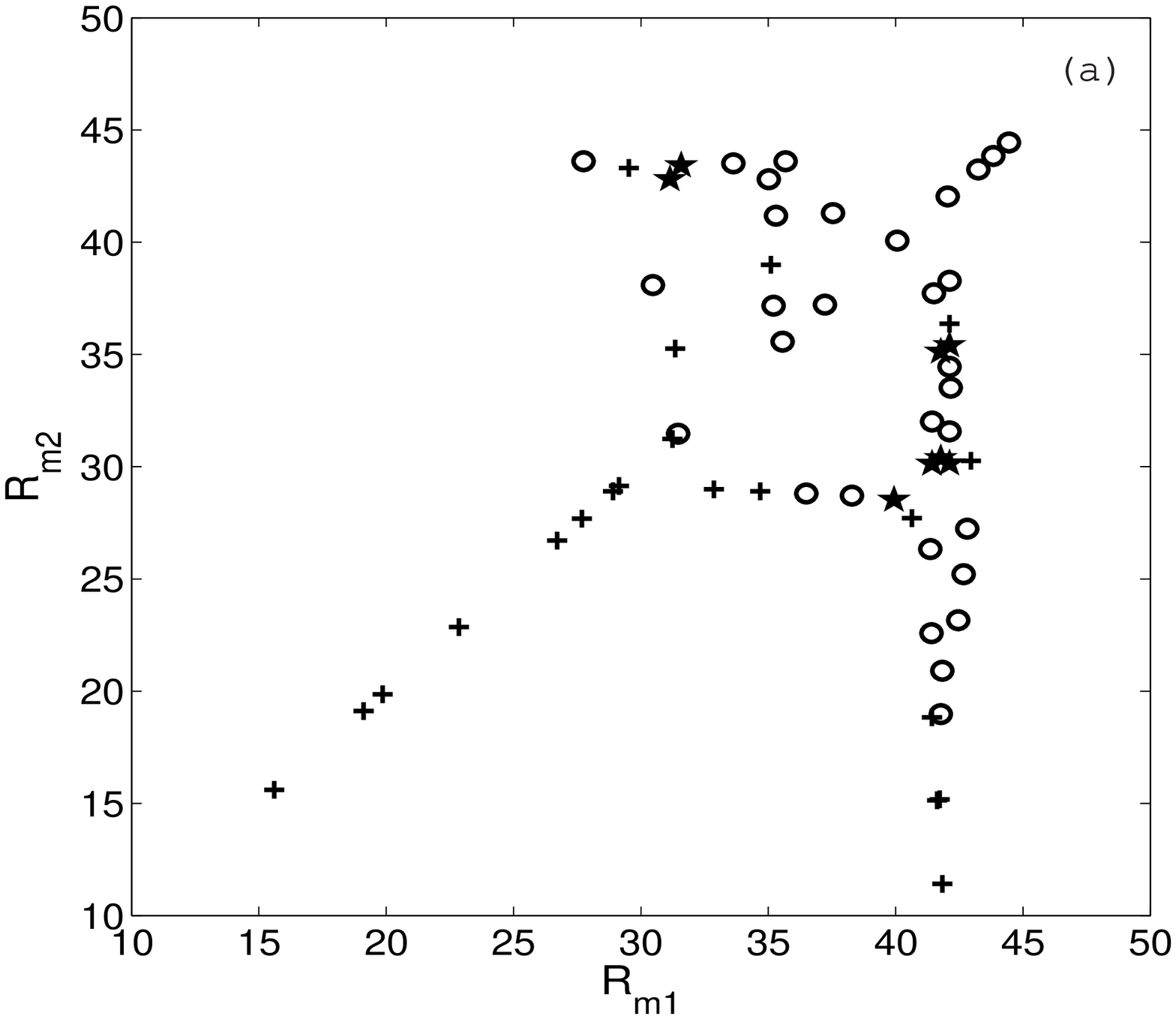}}
\centerline{\includegraphics[width=7cm]{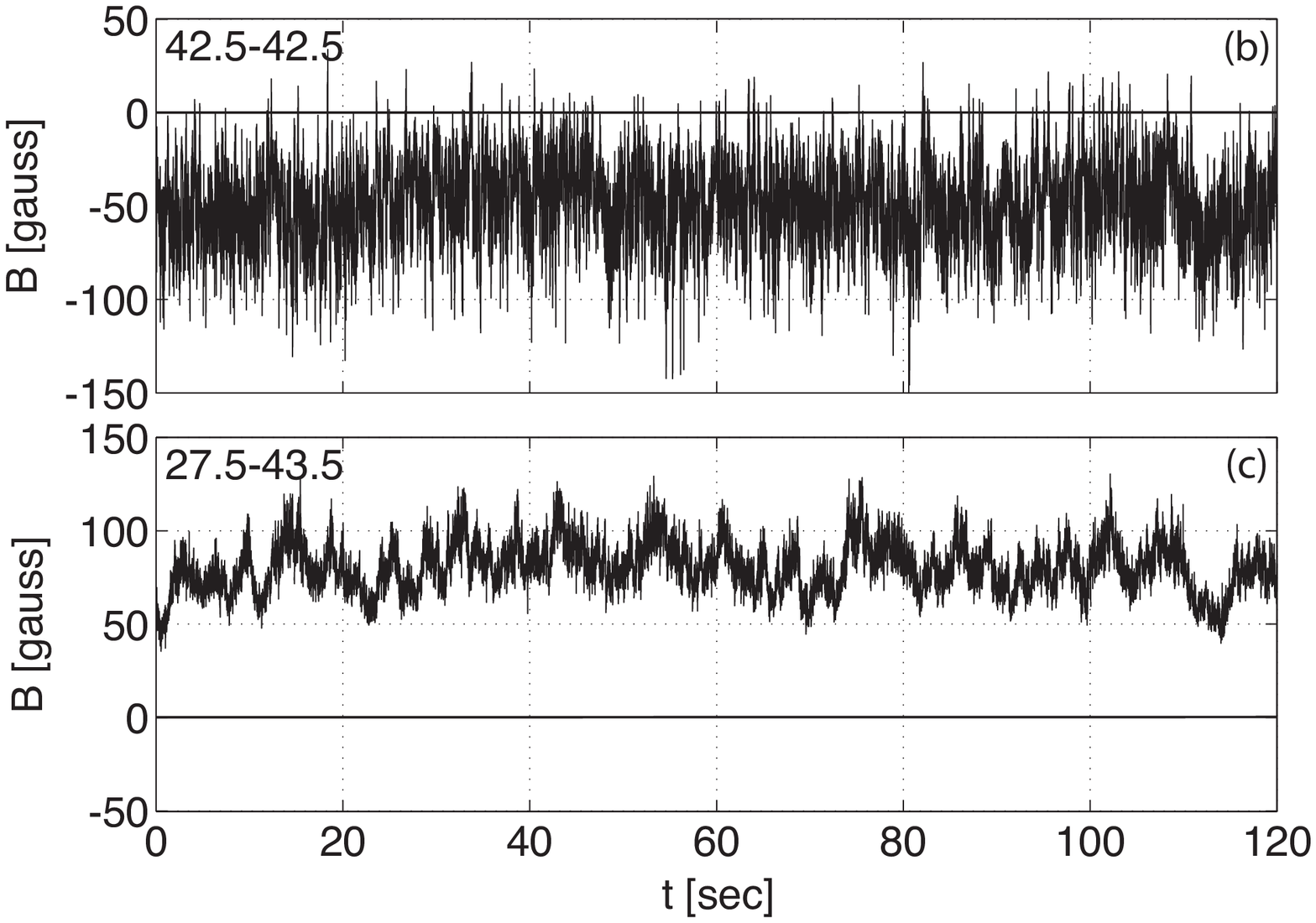}}
\caption{ (a) Preliminary inspection of dynamo regimes observed as the impeller rotation frequencies are independently set. Symbols: $(o)$: statistically stationary dynamos, $(+)$ no dynamo, i.e., magnetic field less than 10~gauss at the measurement location -- and for measurement times longer than 180~s. ($\star$): dynamo with reversals. (b)  Examples of the time variations of the main magnetic field component for rotation frequencies of the disks, $22-22$ Hz (${\rm R_{m1}} = {\rm R_{m2}} = 42.5$) and $14-22$ Hz
(${\rm R_{m1}} = 27.5$, ${\rm R_{m2}} = 43.5)$. Note that on measurement time scales of the order of 180~s, the regimes can depend on the path followed to reach them.
}
\label{param}
\end{figure}

We show in Figure~2 a preliminary inspection of the parameter space accessible when the flow is driven with disks rotating at different speeds.
As said above, only statistically stationary dynamos are observed in the counter-rotating case (Figure 2b). Another statistically stationary dynamo mode is observed when the frequency of one impeller is increased from zero (say $F_1$), the other being kept fixed at $22$ Hz, thus ${\rm R_{m2}}$ in the range $42-43$ depending on the sodium temperature (Figure 2c). Note however that its relative fluctuations are much smaller (compare Figures 2b, c), an effect possibly ascribed to global rotation. This regime undergoes secondary bifurcations when the slower impeller frequency is increased further. In a small parameter range, $\Delta F_i / F_i \approx 20 \%$, a variety of dynamical regimes, oscillations, intermittent bursts (not shown), as well as dynamos with random reversals (Figure 3) are observed. We also find pockets of parameters for which we could not record the growth of a dynamo during 3-minute long runs --corresponding to over 3000 forcing time scales.

\begin{figure*}[!htb]
\begin{center}
\onefigure[width=15cm]
{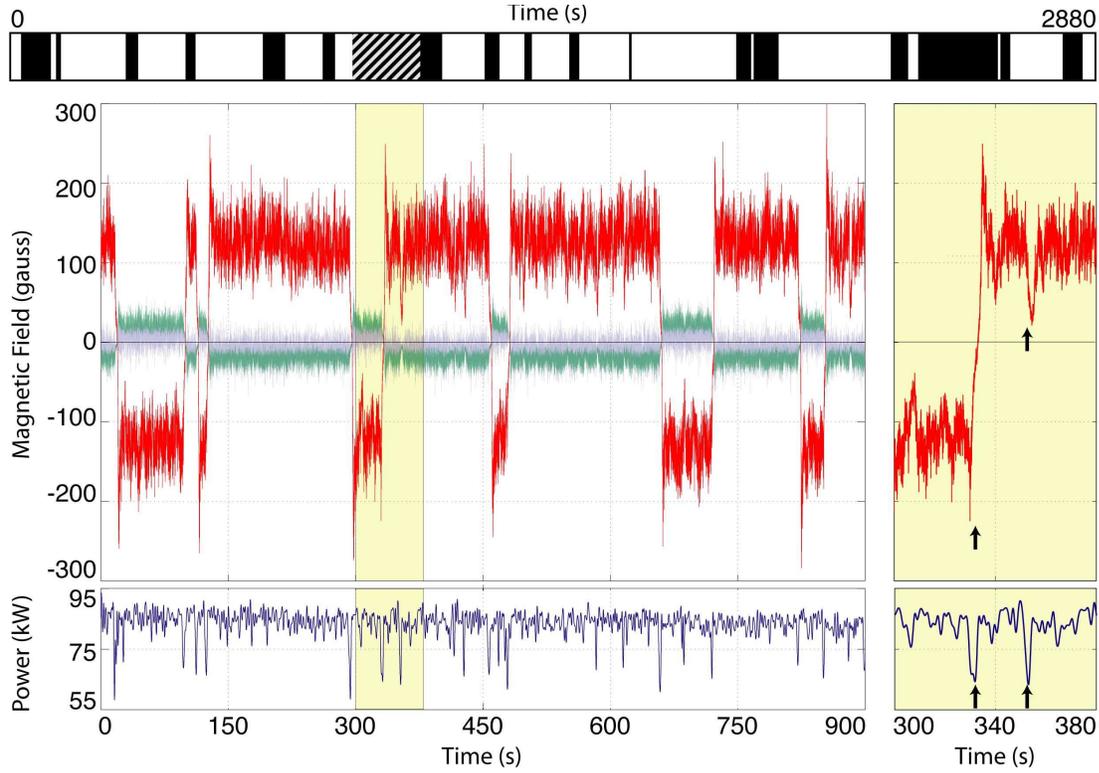}
\caption{Magnetic field measured inside the flow vessel, by a 3-dimensional Hall probe. No external magnetic field is applied, other than the ambient field, whose amplitude is about 0.2 gauss across the measurement volume.The temperature of the outer copper cylinder is  $T=123$ $^{o}$C. (Main): Time evolution of all three magnetic field components. The main component (red) is the azimuthal one. Note that all components decay to zero at a reversal. The bottom graph shows synchronous recordings of the power driving the flow. (Right): detail of the time series of the main magnetic field and simultaneous power consumption (arrows mark the synchronous events). (Top): Chronos of the magnetic field orientation, white for a positive direction, black for the negative direction, for 2 successive recordings 900 and 1800 seconds long (separated by the shaded area, the first sequence corresponds to the main graph). In this regime, the von K\'arm\'an flow is driven with counter-rotating disks at frequencies $F_1=16$~Hz and $F_2=22$~Hz.} 
\label{reverse}
\end{center}
\end{figure*}

We now describe reversals of the magnetic field.
In Figure~3, we show a time series that corresponds to $F_1=16$~Hz and $F_2=22$~Hz. In this regime, the magnetic field reverses at irregular time intervals. All three components of the dynamo field switch polarity in perfect synchrony, so that ${\mathbf B}$ changes to $-{\mathbf B}$. For each polarity, the amplitude of the magnetic field has strong fluctuations, with an rms fluctuation level of the order of 20\% of the mean.  This level of fluctuations is due to the very intense turbulence of the flow, as the kinetic Reynolds number exceeds $10^6$. Reversals occur randomly and have been followed for up to 45 minutes, i.e. 54000 characteristic time scales of the flow forcing.

In the regime reported in Figure~3, the polarities do not have the same probability of observation. Phases with a positive polarity for the largest magnetic field component have on average longer duration ($\langle T_+ \rangle = 120$~s) than phases with the opposite polarity ($\langle T_{-} \rangle = 50$~s). This asymmetry can be due to the ambient magnetic field. Note however that the amplitude of the magnetic field, that is much larger than the Earth field, is the same for both polarities. Standard deviations are of the same order of magnitude as the mean values, although better statistics may be needed to fully converge these estimates. The mean duration of each reversal, $\tau \sim  5$~s, is longer than magnetohydrodynamics time scales: the flow integral time scale is of the order of the inverse of the rotation frequencies, i.e. 0.05~s, and the ohmic diffusive time scale is roughly $\tau_\eta \sim0.4$~s. Concerning the dynamics of field reversals, a natural question is related to the connection between ${\mathbf B}$ and $-{\mathbf B}$ in time. The equations of magnetohydrodynamics are symmetric under the transformation ${\mathbf B}$ to $-{\mathbf B}$ so that the selection of a polarity is a broken symmetry at the dynamo bifurcation threshold. The sequences of opposite polarities act as magnetic domains along the time axis, with Ising-type ÒwallsÓ in-between them: the magnetic field vanishes during the polarity change rather than rotating as in a Bloch-type wall.

One important discovery in these measurements is that reversals of magnetic field are correlated with the global energy budget of the flow. The total power $P(t)$ delivered by the motors driving the flow fluctuates in time in a strongly asymmetric manner: the record shows short periods when $P$ is much smaller than its average. They always coincide with large variations in the magnetic field, as shown in Figure~3. Either a reversal occurs, or the magnetic field first decays and then grows again with its direction unchanged. Similar sequences, called excursions~\cite{valet}, are observed in recordings of the Earth's magnetic field. The variation of power consumption during the weakening of the magnetic field is in agreement with the power required to sustain a steady dynamo in the VKS2 experiment~\cite{monchaux} (drops by over 20\%, that is 20 kW out of 90 kW). However, we note that in other regions of the parameter space, different regimes also involve changes in polarity without noticeable modification of power. 

We have also observed that the trajectories connecting the symmetric states ${\mathbf B}$ and $-{\mathbf B}$ are quite robust despite the strong turbulent fluctuations of the flow. This is displayed in Figure~4: the time evolution of reversals from down to up states can be neatly superimposed by shifting the origin of time such that $B (t=0) = 0$ for each reversal. Despite the asymmetry due to the Earth magnetic field, up-down reversals can be superimposed in a similar way on down-up ones if $-B$ is plotted instead of $B$. For each reversal the amplitude of the field first decays exponentially. A decay rate of roughly $0.8$ s$^{-1}$ is obtained with a log-lin plot (not shown). After changing polarity, the field amplitude increases linearly and then displays an overshoot before reaching its statistically stationary state.\\

\begin{figure}[h!]
\centerline{\includegraphics[width=8cm]{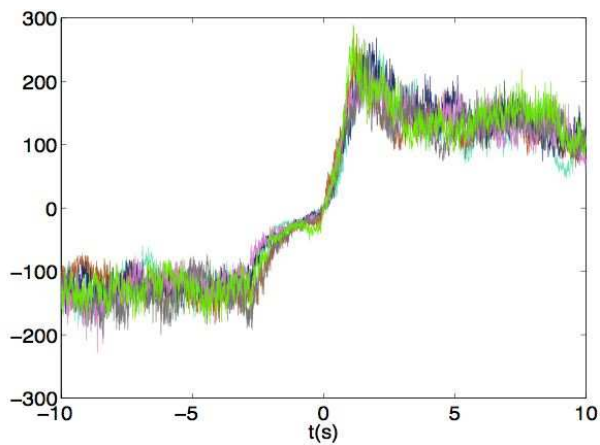}}
\caption{Superimposition of $5$ successive reversals from down to up polarity. For each of them the origin of time has been shifted such that it corresponds to $B = 0$.} 
\label{reverse}
\end{figure}

Further investigation of this regime will help address from an experimental perspective persistent questions about magnetic field reversals. Some of these concern the role of hydrodynamics and electromagnetic boundary conditions -- both of them can be experimentally adjusted. Others are related to the dynamics of the magnetic reversals. From inspection of paleomagnetic data, it has been proposed that reversing dynamos and non-reversing ones are metastable states in close proximity~\cite{macfadden}. In geodynamo simulations (convective dynamos in rapidly rotating spheres), the flow is often laminar and reversals have been associated to interaction between dipole and higher order modes, with the possibility of reversal precursor events~\cite{geophys}.  Fields reversals have also been observed in turbulence driven numerical $\alpha^2$ and $\alpha\omega$ dynamos based on mean-field  magnetohydrodynamics~\cite{alpha}. In these, the role of noise was found to be essential, together with the proximity of steady and oscillating states. In many cases, the existence of several dynamo regimes in a narrow region of parameter space has been considered as essential.  Our experiment displays this feature: two different stationary dynamo modes bifurcate for $F_1=F_2$ and respectively $F_1 \neq F_2$. Their interaction gives rise to a variety of different dynamical regimes in parameter space. This is a general feature for bifurcations of multiple codimension. The most striking aspect of our observation is that the low dimensional dynamics that result from the interaction of a few modes of the magnetic field is preserved despite strong fluctuations of the flow that generates the field. On average, the largest scales of the flow change more than a thousand times during each phase of given polarity. The large scale magnetic field cannot follow turbulent fluctuations and display features characteristic of low dimensional dynamical systems. Flows generating the magnetic fields of planets or stars involve dimensionless parameters (Reynolds and Eckman numbers) orders of magnitudes different from the ones of the present experiment and even further in the case of numerical simulations. However, a weak coupling between the large scale dynamics of their magnetic field and hydrodynamic fluctuations may explain why similar features are observed in some natural dynamos, in numerical modeling and in this experiment.\\

\acknowledgements
We are indebted to Marc Moulin for the technical design of the experiment, and for making so many parts. We thank C\'ecile Gasquet for data acquisition development and for her participation to several campaigns. We thank Jean-Baptiste Luciani and Andr\'e Skiara for their skills in operating the sodium-related equipments. We gratefully acknowledge Didier Courtiade and Jean-Fran\c{c}ois Point for their assistance with the cooling system, Pascal Metz for instrumentation development and Vincent Padilla for making parts of the experiment. We thank the ``Dynamo'' GDR 2060, and numerous colleagues with whom we have had fruitful discussions over the years, in particular Emmanuel Dormy for useful discussions about the characteristics of the dynamics of the Earth magnetic field. This work is supported by the French institutions: Direction des Sciences de la Mati\`ere and Direction de l'Energie Nucl\'eaire of CEA, Minist\`ere de la Recherche and Centre National de Recherche Scientifique (ANR 05-0268-03). The experiment is operated at CEA/Cadarache Ð DEN/DTN.\\

\noindent {\it present addresses:} \\
(L.M., +) IFREMER / Laboratoire de Physique des Oc\'eans, CNRS~UMR~6523,  BP70,  F-29280,  Plouzane (France)\\
(F.R. *) Laboratory for Aero and Hydrodynamics, TU-Delft (The Netherlands)\\
(M.B. \#) Laboratoire des \'Ecoulements G\'eophysiques et Industriels, CNRS UMR 5519,  BP53, F-38041 Grenoble (France)\\

\end{document}